\documentclass[twocolumn]{aastex631}
\usepackage{appendix}
\usepackage{listings}

\usepackage{savesym} 
\savesymbol{tablenum}
\usepackage{siunitx}
\restoresymbol{SIX}{tablenum}
\usepackage{siunitx}
\usepackage{amsmath, amssymb} 
\usepackage[ruled,vlined]{algorithm2e}
\usepackage{hyperref} 
\usepackage{nameref} 
\usepackage{amsfonts}      
\usepackage{listings}
\usepackage{svg}
\usepackage{booktabs}
\usepackage{xcolor}
\usepackage{natbib}
\usepackage{bbm}
\usepackage[utf8]{inputenc} 

\begin{document}

\title{Practical Author Name Disambiguation under Metadata Constraints: A Contrastive Learning Approach for Astronomy Literature}

\keywords{Author Name Disambiguation, Machine Learning, Open Science, Digital Libraries}













\author[0000-0002-2248-0941]{Vicente Amado Olivo}
\affiliation{Department of Computational Mathematics, Science, and Engineering, Michigan State University, East Lansing, MI 48824, USA}

\author[0000-0002-0479-7235]{Wolfgang Kerzendorf}
\affiliation{Department of Computational Mathematics, Science, and Engineering, Michigan State University, East Lansing, MI 48824, USA}
\affiliation{Department of Physics and Astronomy, Michigan State University, East Lansing, MI 48824, USA}

\author{Bangjing Lu}
\affiliation{Department of Computational Mathematics, Science, and Engineering, Michigan State University, East Lansing, MI 48824, USA}

\author[0000-0002-1560-5286]{Joshua V. Shields}
\affiliation{Department of Physics and Astronomy, Michigan State University, East Lansing, MI 48824, USA}

\author{Andreas Flörs}
\affiliation{GSI Helmholtzzentrum für Schwerionenforschung, Darmstadt, Germany}

\author[0000-0001-7571-0742]{Nutan Chen}
\affiliation{Foundation Future Industries, Munich, Germany}

\correspondingauthor{Vicente Amado Olivo}
\email{amadovic@msu.edu}

\begin{abstract}
The ability to distinctly and properly collate an individual researcher's publications is crucial for ensuring appropriate recognition, guiding the allocation of research funding, and informing hiring decisions. However, accurately grouping and linking a researcher's entire body of work with their individual identity is challenging because of widespread name ambiguity across the growing literature. Algorithmic author name disambiguation provides a scalable approach to disambiguating author identities, yet existing methods have limitations. Many modern author name disambiguation methods rely on comprehensive metadata features such as email, venue, or affiliation. Despite advancements in digitally indexing publications, metadata is often unavailable or inconsistent in large digital libraries (e.g., SAO/NASA Astrophysics Data System). We introduce the Neural Author Name Disambiguator, a method that effectively disambiguates author identities in large digital libraries despite limited metadata availability. We formulate the author name disambiguation task as a similarity learning problem by employing a Siamese neural network to disambiguate author names across publications relying solely on widely available publication metadata—author names, titles, and abstracts. We construct the Large-Scale Physics ORCiD-Linked dataset to evaluate the Neural Author Name Disambiguator method by cross-matching NASA/ADS publications ORCiD. By leveraging foundation models to embed metadata into features, the Neural Author Name Disambiguator achieves up to 94\,\% accuracy in pairwise disambiguation and over 95\,\% F1 in clustering publications into their correct researcher identities. We release the testing dataset as a benchmark for physics and astronomy, providing realistic evaluation conditions for future disambiguation methods. The Neural Author Name Disambiguator algorithm demonstrates effective disambiguation with minimal metadata, offering a scalable solution for name ambiguity in large-scale digital libraries.
\end{abstract}

\section{Introduction}\label{sec:intro} 
Consolidating a researcher’s publications into a unified collection is essential for proper credit, funding allocation, hiring, and downstream tasks such as expertise identification \citep[see, e.g.,][]{kurtz_effect_2005, eggert_best_2011, portenoy_bursting_2022, strolger_pacman2_2023}. The exponential growth of the scientific literature and the global number of scientists has exacerbated the challenge of tracking an individual’s complete body of work \citep{noauthor_estimates_nodate, kerzendorf_knowledge_2019, schneegans_unesco_2021, schneider_publications_2023}. The task of linking publications to the correct researcher is especially challenging as authors share names or appear under multiple name variations. In some disciplines up to two-thirds of authors are affected by ambiguous names \citep{torvik_author_2009}. 

Consequently, search algorithms embedded in digital libraries merge distinct authors under the same name. For example, the SAO/NASA Astrophysics Data System (NASA/ADS)\footnote{\url{https://ui.adsabs.harvard.edu/}} returns over \num{20000} records for J.~Smith and \num{280000} records for Y.~Wang, illustrating the ambiguity caused by common names. The Open Researcher and Contributor ID (ORCID) was introduced as a solution to uniquely link publications to the correct researcher. However, researcher enrollment in ORCiD remains limited with \citet[][]{bello_charting_2020} estimating an adoption rate of only 60\% across fields. Algorithmic solutions are still essential for historical records and publications without identifier coverage. Thus, Author Name Disambiguation (AND) remains a foundational task for libraries to ensure accurate attribution of scholarly work \citep[see, e.g.,][]{borgman_gettys_nodate}.

Advances in machine learning have facilitated many sophisticated solutions for AND. Most state-of-the-art AND algorithms rely on supervised learning methods and require extensive metadata for each publication \citep[such as affiliations, coauthor names, or titles;][]{pooja_exploiting_2022, liu_author_2024}. In practice, however, such metadata is often unavailable, inconsistent, or erroneous in large digital libraries \citep{choudhury_metaenhance_2023} . For example, we find nearly 30\% of NASA/ADS entries are missing affiliations. In the absence of reliable metadata, many AND methods are evaluated on manually labeled benchmark datasets that have clean, complete, and curated records \citep{kim_model_2020}. Incomplete records are often discarded, which can artificially inflate metrics such as recall \citep{gurney_author_2011}. \citet{zhou_evaluating_2024} find that missing affiliations degrade large-scale disambiguation and introduce systematic anomalies. In summary, disambiguation models perform well on cleaned benchmark datasets but struggle to generalize when data distributions shift in real-world applications \citep{kim_model_2020, koh_wilds_2021}.

To address these challenges, we introduce the Neural Author Name Disambiguator (NAND), a zero-shot similarity learning approach based on contrastive training. NAND learns to distinguish between matching and non-matching author publications using only the most common metadata (i.e., author name, titles, and abstracts). NAND avoids dependence on richer but often unreliable metadata such as affiliations or coauthor networks by relying on minimal and widely available features. Alongside the model, we release the Large-Scale Physics ORCiD-Linked (LSPO) dataset, sampled from NASA/ADS publications, which provides the first large-scale ORCiD-linked benchmark for real-world disambiguation in physics and astronomy. We share this dataset to allow benchmarking for real-world applications. 

The paper is organized as follows: In Section~\ref{sec:relwork}, we review prior work on author name disambiguation, including  methodological approaches and dataset construction. Section~\ref{sec:data} describes the AND dataset we constructed for physics and astronomy using NASA/ADS. Section~\ref{sec:methods} details the NAND algorithm, highlighting its reliance on minimal metadata features. Section~\ref{sec:results} presents our evaluation across datasets, examining both performance within astronomy and generalizability to other domains. Section~\ref{sec:discussion} discusses the comparison of prior methods and broader implications. We conclude this paper in Section~\ref{sec:conclusion}.  

\begin{table*}[t!]
\centering
\caption{Comparison of datasets for Author Name Disambiguation. Values in parentheses refer to the available records in physics.}
\label{tab:AND_comparison}
\setlength{\tabcolsep}{5pt}
\resizebox{\textwidth}{!}{%
\begin{tabular}{l c r r c}
\toprule
\textbf{Dataset} & \textbf{Domain/Focus} & \textbf{Publications} & \textbf{Researchers} & \textbf{Limitations} \\
\midrule
LAGOS-AND \citep{zhang_span_2023} & Various domains & 7,510,203 (222,517) & 797,901 (69,475) & Not suited for domain-specific tasks\\
AUT-ORC \citep{kim_orcid-linked_2021} & Biomedicine & 3,076,501 & 245,755 & Domain limited \\
S2AND \citep{subramanian_s2and_2021} & Various domains & 487,567 & -- & Based on past datasets \\
INSPIREHEP \citep{ngonga_ngomo_ethnicity_2016} & High energy physics & 360,066 & 15,388 & Domain limited \\
Aminer \citep{zhang_name_2018} & Various domains & 70,258 & 12,798 & Focused on Chinese names \\
SCAD-ZbMath \citep{muller_data_2017} & Mathematics & 33,810 & -- & Domain limited \\
NASA/ADS \citep{mihaljevic_disambiguation_2021} & Physics and astrophysics & 9,545 & 562 & Small, Domain limited \\
Arnetminer \citep{wang_adana_2011} & Various domains & 6,730 & 100 & Small \\
PubMed \citep{song_exploring_2015} & Biomedicine & 2,875 & 385 & Small, Domain limited \\
QIAN (DBLP) \citep{noauthor_dynamic_nodate} & Computer science & 1,322 & 171 & Domain limited \\
Medline \citep{vishnyakova_new_2019} & Biomedicine & 1,900 pairs & -- & Domain limited \\
LSPO & Physics & 553,496 & 125,486 & Domain limited \\
\bottomrule
\end{tabular}%
}
\end{table*}

\section{Related Work}\label{sec:relwork}
We present an overview of the existing methodologies and datasets addressing the AND problem. AND solutions span heuristic, supervised, unsupervised, and graph-based methods. The available datasets for training and testing disambiguation methods cover a range of fields, vary in size, and originate from different digital libraries.
\subsection{AND Methods}
Heuristic-based methods apply matching rules to author names and metadata. For example, name-matching rules compare first and middle initials or combine heuristics with metadata such as coauthor names or affiliations \citep{newman_2001, strotmann_author_2010, dangelo_heuristic_2011, wang_author_2011,  milojevic_accuracy_2013}. While simple and interpretable, these approaches are sensitive to incomplete or erroneous data and do not scale efficiently to large digital libraries \citep{subramanian_s2and_2021}. 

Supervised methods leverage labeled datasets to learn patterns distinguishing authors by learning features from names, coauthors, affiliations, and other metadata \citep{han_two_2004, tran_author_2014, boukhers_deep_2023}. For instance, \citet{mihaljevic_disambiguation_2021} trained a random forest classifier on author names, coauthors, and abstracts from a small manually labeled NASA/ADS dataset. \citet{tran_author_2014} developed a deep neural network for Vietnamese names, where textual features (i.e., names, affiliations, coauthors, keywords) were represented with TF--IDF and Jaccard similarity. \citet{boukhers_deep_2023} combined semantic representations of textual metadata with symbolic representations of author and coauthor names to disambiguate records in the DBLP\footnote{\url{https://dblp.org/}} digital library. Supervised methods perform well when labeled data are available, but compiling representative training datasets at scale remains challenging, limiting their effectiveness \citep[see, e.g.,][]{kim_evaluating_2018, zhou_evaluating_2024, subramanian_s2and_2021}.  

Unsupervised methods aim to cluster author records without labels, often exploiting graph representations, similarity measures, or clustering \citep{ferreira_effective_2010}. \citet{zhang_name_2018}, for example, introduced a representation learning framework in Aminer that combined global similarity learning with a local linkage graph refined by a graph autoencoder. \citet{khabsa_large_2014} applied density-based clustering with a random forest distance function, later extended with temporal and name constraints \citep{ester_density-based_1996, breiman_random_2001, khabsa_online_2015}. Unsupervised methods are flexible across domains and can exploit large unlabeled datasets, but can propagate errors if the underlying data are erroneous \citep[see, e.g.,][]{choudhury_metaenhance_2023, zhou_evaluating_2024}. 

Graph-based methods model authors, publications, and affiliations as networks to capture relationships across records \citep{fan_graph-based_2011, qiao_unsupervised_2019, xiong_learning_2021}. \citet{santini_knowledge_2022} presented LAND, which uses multimodal graph embeddings combined with heuristic features. \citet{pooja_exploiting_2022} proposed an attention-based multi-dimensional graph convolution network that captures higher-order document relationships. \citet{liu_author_2024} introduced the ARCC framework, which refines paper association graphs and applies compositional contrastive learning to produce discriminative representations. Despite these advances, graph-based methods remain sensitive to noise or missing edges, which can cause inaccuracies to spread in large incomplete bibliographic datasets \citep{liu_author_2024}.

\subsection{AND datasets}\label{sec:datasets}
Representative and reliable labeled datasets are essential both for training supervised algorithms and for evaluating disambiguation methods \citep{subramanian_s2and_2021}. AND datasets vary in scale and domain, from small manually-labeled datasets to large semi-automatic resources. An early dataset from the Arnetminer digital library \citep{wang_adana_2011} annotated \num{6730} published papers from a set of \num{100} highly ambiguous author names. The second generation digital library Aminer \citep{zhang_name_2018} extends this dataset by compiling a benchmark of \num{70258} documents from \num{12798} authors, with a focus on \num{100} primarily Chinese names. The WhoIsWho benchmark further extends this to over \num{1000000} papers from Aminer, linking \num{70000} authors sampled from \num{1000} unique names \citep{chen_web-scale_2023}.

In the computer science domain, \citet{noauthor_dynamic_nodate} presented a dataset of \num{1322} author records for \num{171} researchers in the DBLP digital library. While in the biomedical field, the PubMed digital library \citep{song_exploring_2015} constructed a gold standard dataset of \num{385} authors and \num{2875} publications. Similarly, in the domain of mathematics, \citet{muller_data_2017} compiled a dataset of approximately \num{30000} author records from the DBLP and ZbMath digital libraries. Furthermore, in astrophysics, \citet{mihaljevic_disambiguation_2021} presented a manually labeled dataset from the NASA/ADS database of \num{9545} author records from \num{562} researchers.

Large-scale semi-automatic datasets aim to improve coverage and diversity. The SCAD-SbMath dataset \citep{muller_data_2017} presented manual and automatic disambiguation for dataset creation through co-author graphs. The Inspire-HEP digital library combined crowd-sourcing and expert curation to compile a dataset of roughly \num{15388} researchers across \num{360066} publications \citep{ngonga_ngomo_ethnicity_2016}. 

Cross-dataset integration and ORCID-linked publication data have enabled large-scale benchmarks. S2AND combined records from multiple prior datasets to create a diverse dataset covering many domains, representing approximately 82\% of the combined datasets \citep{subramanian_s2and_2021}. ORCID-based datasets include linking Medline to ORCID, which produced \num{245755} unique authors in biomedical research \citep{kim_orcid-linked_2021}. LAGOS-AND links the OpenAlex library to ORCID resulting in \num{797901} unique authors across scientific domains \citep{zhang_span_2023}. Table~\ref{tab:AND_comparison} summarizes existing resources and the LSPO dataset.

\section{Data}\label{sec:data} 
In this section, we describe the LSPO dataset \footnote{\url{https://zenodo.org/records/11489161}}  developed for training and evaluating the NAND model. We detail the data sources, cross-matching to ORCiD, and the negative sampling strategy for contrastive learning. LSPO addresses the lack of large-scale datasets in physics and astronomy for the AND task, where existing resources are either small or limited to subfields. We release LSPO alongside NAND as a benchmark for author name disambiguation in astronomy. To evaluate generalizability beyond physics and astronomy, we also compare to LAGOS-AND, a large-scale ORCiD-linked dataset spanning multiple scientific domains.

\begin{figure*}[t!]
\includegraphics[width=\textwidth]{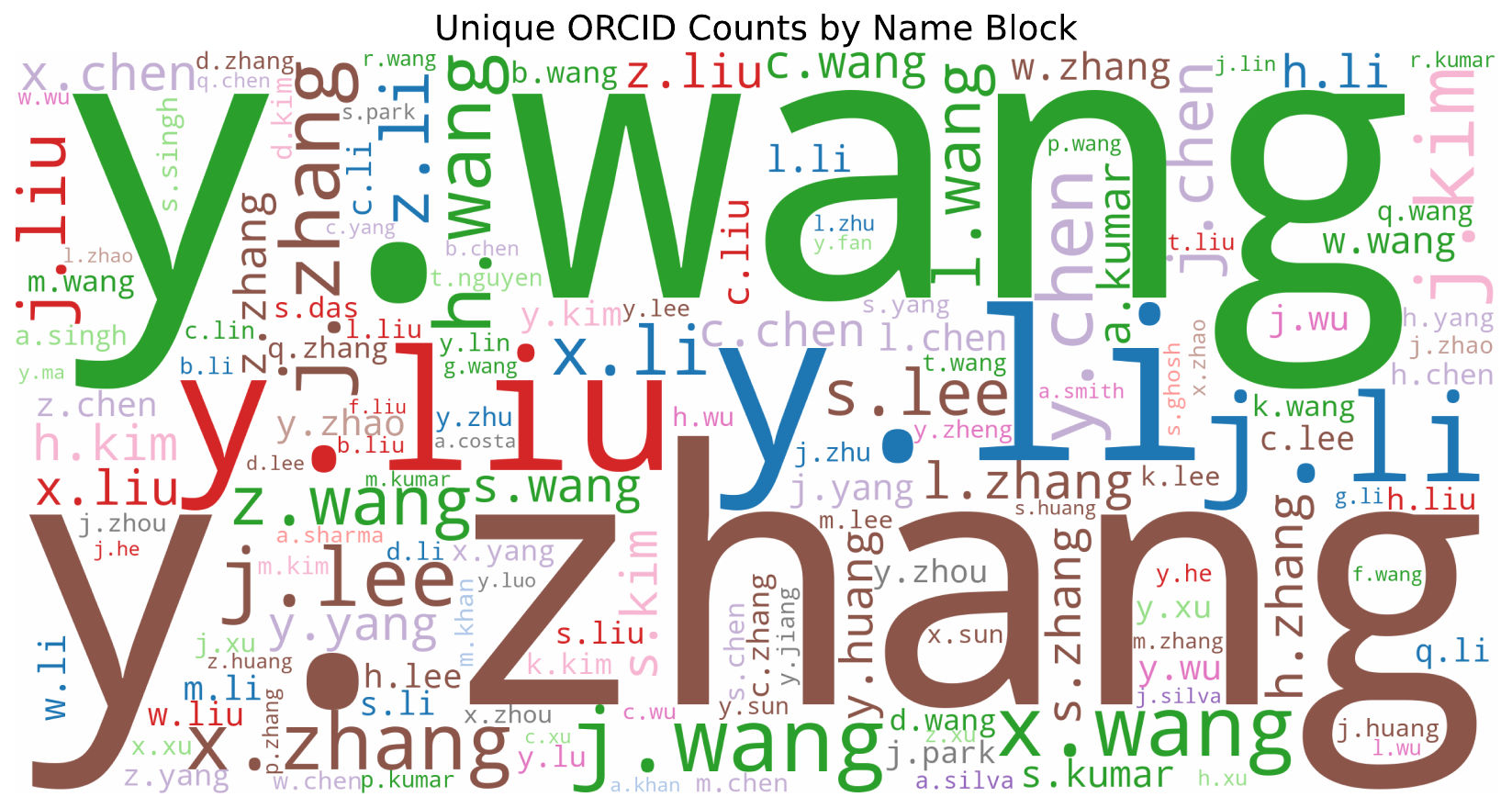}
\caption{Top 150 name blocks with the highest number of unique ORCID identifiers in our dataset, represented by word size. One of the largest name blocks, Y. Wang, contains 153 unique ORCID identifiers across 426 author record instances. In contrast, the majority of the 97,936 name blocks (88,369) are associated with a single unique ORCID identifier.}
\label{fig:bubble_chart}
\end{figure*}

\subsection{LSPO Dataset Construction}
The LSPO dataset consists of author records, where each record links an author’s name to a specific publication. We extract publication data from NASA/ADS, filtering for research works up to 2021, totaling \num{16539968}. The NASA/ADS titles are then cross-matched with ORCiD profile data up to the year 2020 to provide labels \citep{Montenegro_2023}. Assigning ORCiD identifiers to authors in NASA/ADS publications requires strict matching criteria, as author names on publications do not always align with their names on ORCiD profiles. Variations in author name formatting include different combinations of initials and full names (e.g., first initial and last name or first and middle initials with last name). We first check for an exact match between the author name on NASA/ADS and the ORCiD profile to ensure accurate matches. If an exact match is found, we assign the ORCiD to that author. For cases with only a partial match (e.g., first initial and last name), we verify that no other author on the publication shares the same partial match before assigning the ORCiD. If multiple authors on the same paper share the same first initial and last name, we exclude the ORCiD to prevent erroneous assignments across the dataset. The resulting dataset after cross-matching and ensuring accurate matches is \num{553496} author records from NASA/ADS linked to \num{125486} unique ORCiD identifers. The available metadata span textual components such as title and abstract, unique identifiers such as arXiv identifiers and DOIs, and researcher information (i.e., author names or affiliation). 


\subsection{LSPO statistics}
The LSPO dataset is sampled directly from NASA/ADS, ensuring it reflects the true distribution of AND challenges in physics and astronomy. Name formatting in LSPO mirrors the variability in NASA/ADS, with 42\% of first names appearing as initials, 37\% as full first name, and the remainder in more complex forms (e.g., middle initials or multiple first initials). To organize the disambiguation task, we group author records into name blocks, where all records share the same first initial and last name (e.g., Y.~Zhang or D.~Torres). LSPO contains 97,936 name blocks. For example, the block Y.~Zhang includes 555 records linked to 143 unique ORCiDs, while D.~Torres has 407 records split between only two ORCiDs. Figure~\ref{fig:bubble_chart} illustrates the top 150 name blocks with the highest number of unique ORCID identifiers. The NAND model is trained on widely available metadata fields found across large digital libraries: author names, titles, and abstracts. Figure~\ref{tab:metadata_comparison} presents the coverage of different metadata in the LSPO dataset. We experiment with including different metadata fields as features in our ablation studies, such as affiliations and publication venue. However, affiliations and other attributes are often erroneous or inconsistently available.
We compare LSPO with a curated NASA/ADS corpus of over 10 million refereed physics and astronomy articles. In both datasets, more than 40\% of author first names appear as initials only, while in benchmark datasets such as Aminer-18 this rate is below 5\%. Prior work has shown that the distribution of publications per author is highly skewed, with most authors having few publications \citep{a_j_lotka_frequency_1926, yan_knowledge-based_2023}. Similarly, over 42\% of name blocks in both LSPO and NASA/ADS contain only a single author record (e.g., the A. Abdelaziz name block has just one publication). Metadata availability is similar across LSPO and NASA/ADS. However, NASA/ADS provides abstracts for only about 60\% of records; we raise this to over 94\% by augmenting with data from the OpenAlex digital library \citep{priem_openalex_2022}. For training and evaluation, we construct balanced pairs of author records. Pairs are labeled positive if both records share an ORCiD and negative otherwise. Our pair construction emphasizes the most challenging ambiguities to yield a realistic large-scale benchmark for AND.

\subsection{Pairs and negative sampling}
To train an effective AND method, it is crucial to carefully construct the training data by identifying relevant and ambiguous pairs of author records. Pairing all \num{553496} NASA/ADS works linked to ORCiD, an exhaustive pairwise comparison, results in $\approx 10^{11}$ pairs for training, validation, and testing. However, a significant portion of these pairs can be simply disambiguated as the first initial and last names of the two author records do not match. We group author records into name blocks to focus on the most challenging and ambiguous cases.

We first divide author records by ORCiD identifiers to prevent data leakage, where information from the test set could influence training. We allocate 60\% of ORCiD identifiers to training, 20\% to validation, and 20\% to testing, ensuring that no data from the same ORCiD appears in more than one split before constructing pairs. Additionally, to avoid introducing noise during training and evaluation, we omit pairs of author records from the same publication. For instance, large author lists on physics publications may contain multiple authors within the same name block but with distinct ORCiD identifiers.

The most difficult cases to disambiguate occur when individuals share identical or very similar names. We capture these challenges by constructing hard negative pairs, records from the same name block that belong to different researchers, within each data split. In total, the dataset contains \num{9618807} positive and hard negative pairs derived from author records in the same name block. However, the \num{9618807} pairs are unbalanced across the two classes, with \num{8269418} positive pairs and \num{1349389} hard negative pairs. Therefore, we balance the dataset by downsampling to obtain an equal number of positive and hard negative pairs, resulting in a total of \num{2698778} pairs for training and evaluation. The number of training pairs is \num{2036344}, validation pairs is \num{268668}, and testing pairs is \num{393766}. The validation and testing sets are unequal because we first split by ORCiD to prevent data leakage, which results in differing numbers of pairs depending on which ORCiDs fall into each data split.	

\subsection{LSPO limitations}
It is important to note that the comprehensiveness of our dataset is constrained by the coverage of researchers who have adopted and updated their ORCiD profiles with their published works. Additionally, when cross-matching publications from NASA/ADS with ORCiD profiles, we applied a filtering process to exclude potentially mismatched or generic titles present in NASA/ADS and other digital libraries (e.g., "Letter to the Editor" or "news and notes"; see Appendix~\ref{app:processing} for the specific query used in the filtering process). Furthermore, some researchers have multiple ORCiD identifiers associated with their publications, and some ORCiD profiles contain duplicate entries of a research work with mismatching metadata. We removed such ORCiD profiles from our dataset to avoid biasing our model. Initially, we had a total of \num{126321} ORCiD identifiers, but removed \num{835} ORCiD identifiers. 

\subsection{Data Comparison}
We assess the performance of our NAND method beyond physics and astronomy by comparing against the large-scale, multidisciplinary LAGOS-AND dataset \citep{zhang_span_2023}. LAGOS-AND is an ORCID-linked dataset spanning a wide range of scientific domains by cross-matching ORCiD and the OpenAlex digital library. We therefore train and evaluate NAND on LSPO and compare it with a model trained on LAGOS-AND, measuring both in-domain performance and cross-domain transfer.

\begin{figure*}[t!]
\includegraphics[width=\textwidth]{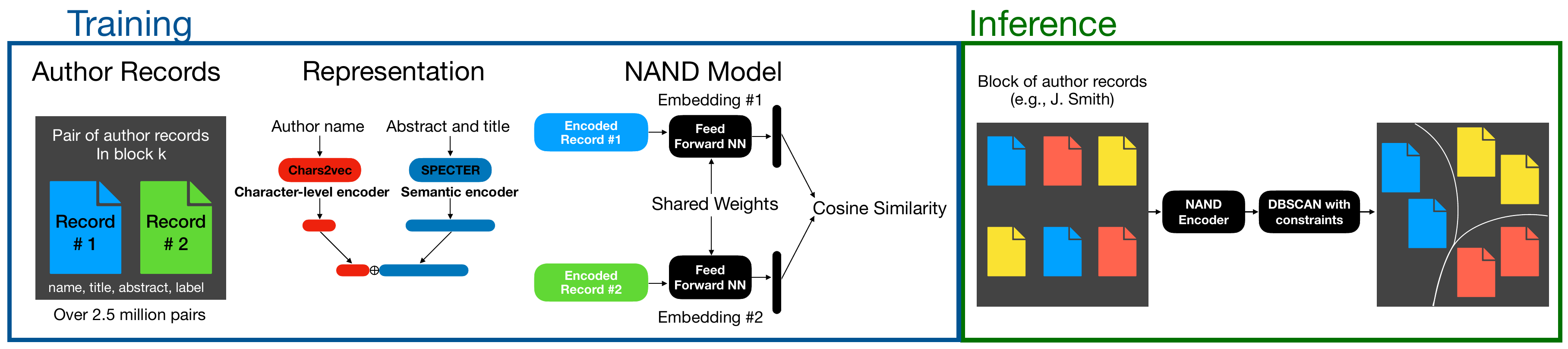}
\caption{Overview of the pairwise similarity model training process and the clustering method: (1) Representing researcher information (e.g., author name) and publication content (e.g., title and abstract) using foundation models. (2) Training a Siamese neural network on the represented author records to distinguish if two author records refer to the same researcher or different researchers. (3) Clustering full name blocks of represented and transformed embeddings of author records utilizing DBSCAN with added constraints for naming conventions and years in-between publications.}
\label{fig:AND_framework}
\end{figure*}

\section{Methods} \label{sec:methods}
In this section, we introduce the Neural Author Name Disambiguator algorithm (NAND) for disambiguating pairs of author records. We also present the representation methodologies and loss functions considered for training.

\subsection{Problem Formulation}

We design NAND to address the task of AND by framing it as a similarity learning problem. Let 
\(\mathcal{S} = \{s_1, s_2, \dots, s_n\}\) denote a set of publications, and let 
\(\mathcal{A} = \{A_1, A_2, \dots, A_n\}\) be the corresponding sets of author names for each publication, where 
\(A_i = \{a_{i1}, a_{i2}, \dots, a_{im_i}\}\) contains all authors of publication \(s_i\).  

We define an author record \(r_{i}\) as a single author \(a_{ij}\) together associated with a specific publication and its metadata (e.g., author, title, abstract, and affiliation). The set of all author records across the dataset is then  

\[
\mathcal{R}= \bigcup_{i=1}^n A_i = \{ r_1, r_2, \dots, r_N \},
\]  

where \(N = \sum_{i=1}^n |A_i|\) is the total number of author records, and \(|A_i|\) denotes the number of authors in publication \(s_i\).

To reduce the search space for disambiguation, author records are grouped into name blocks, where each block contains all author records sharing the same first initial and last name. Formally, let 
\(\mathcal{B} = \{B_1, B_2, \dots, B_n\}\) denote the set of name blocks. Each name block \(B_k\) contains a subset of author records that share the same first initial and last name, such that
\[
B_k \subset \mathcal{R},\, 
\bigcup_{k=1}^n B_k = \mathcal{R}, \, 
B_i \cap B_j = \emptyset \, \forall i,j \in \{1,\dots,n\}, \,i \neq j.
\]

In other words, each author record belongs to exactly one name block, and together the blocks cover all records in the dataset without overlap. Within each block \(B_k\), the disambiguation task reduces to partitioning the author records into disjoint clusters  
\[
B_k = C_{k1} \cup C_{k2} \cup \dots \cup C_{k\ell_k}, \quad 
C_{ki} \cap C_{kj} = \emptyset \text{ for } i \neq j,
\]  
where each cluster \(C_{ki}\) contains all publications written by a single distinct researcher. We perform this partitioning via clustering in the learned embedding space of the NAND model. 

The partitioning is performed by embedding the records, \(r_{i}\), and associated metadata in a new space using a function $f =$ \(\phi(\cdot; \theta)\), to then compute the similarity of the two embedded publication metadata vectors using $\mathrm{sim}(\cdot, \cdot)$, any similarity function between embeddings. Finally, clustering is applied within each name block in the learned embedding space \(Z\) containing the learned embeddings \(z_{i}\). Figure~\ref{fig:AND_framework} illustrates the full NAND pipeline of author record representation, contrastive learning, and clustering into disambiguated researcher identities.

\subsection{Author Record Representation}
We utilize different methods to represent the information of each researcher (i.e. author name) and publication content (i.e. title and abstract) of each author record for training the NAND model. For researcher information, we test the chars2vec character-level representation method \citep{cao_joint_2016} to handle variations in string placement, such as flipped author name order or parsing errors. The chars2vec model embeds similar words close to each other in the embedding space, regardless of where characters appear in the word (e.g., misspellings, character order), making it robust to variations. For comparison, we evaluate the word2vec model \citep{mikolov_efficient_2013} on entire names to assess how character-level or word-level information impacts performance. For publication content, we utilize transformer-based language models that can embed similar scientific documents close together in the embedding space. The \textsc{Specter} model \citep{cohan_specter_2020} captures semantic and contextual information in scientific texts and was developed for the similarity of scientific documents. In addition to the \textsc{Specter} model, we test the Llama 3.2 model \citep{grattafiori_llama_2024} to evaluate the benefits of general embeddings to maximize disambiguation performance. To output embeddings, we utilize the CLS token for the BERT-based \textsc{Specter} model, and mean pooling for Llama 3.2. We set the maximum input length to 256 tokens for \textsc{Specter} and Llama 3.2.

\subsection{Contrastive Learning}
We employ a Siamese neural network architecture for our AND similarity learning task. The Siamese network consists of two identical subnetworks that share the same parameters and weights. Each subnetwork takes an author record $r_i$ as input and maps it to a dense embedding vector $z_i$ in a latent space:

\begin{equation}
z_i = \phi(r_i; \theta)
\end{equation}

where $\phi$ is the embedding function parameterized by $\theta$, which is learned during training.

The objective is to learn an embedding space where the distance between embeddings of author records belonging to the same author is smaller, while the distance between embeddings of author records belonging to different authors is larger.
The embedding function, $\phi$, incorporates both the content of publications, which is measured by the similarity in their embeddings, and the author labels. The dual approach ensures that records from the same author are grouped closer together, while records from different authors are separated in the embedding space. The model is trained once on LSPO ORCiD-linked pairs and can be applied to entirely new name blocks without retraining. This enables zero-shot disambiguation, where the method generalizes to unseen authors at inference.

In this work, we evaluate two different loss functions to train the Siamese network and update the learned embeddings: the cosine embedding loss and the Noise-Contrastive Estimation InfoLoss (InfoNCE). 

The cosine embedding loss minimizes the cosine distance of a pair of samples from the same class but maximizes the cosine distance of samples from different classes. For a pair of embedded author records $(z_i, z_j)$ and a label $y$, the cosine-embedding loss \citep{Pal2017CopernicanL} is defined as:

\begin{equation}\label{eq:2}
\ell_{i,j} = 
\begin{cases} 
1 - d(z_i, z_j) & \text{if } y = 1 \\
\max(0, d(z_i, z_j) - \text{margin}) & \text{if } y = -1 
\end{cases}
\end{equation}

Where $d(z_i, z_j)$ denotes the distance (e.g., cosine distance) between the dense embedding vectors $(z_i, z_j)$ in the latent space, and $y$ denotes whether a pair of author records belongs to the same researcher (i.e., $y=1$) or different researchers (i.e., $y=-1$). Specifically, the cosine-embedding loss measures the similarity between a pair of samples by computing the cosine similarity between the embeddings in the space.

The InfoNCE Loss calculates the similarity between a pair of sample representations $(z_i, z_j)$, and contrasts it with the similarity of the representations to other representations within the batch of samples. This is calculated by the sum of the exponential similarity scores between the representation $z_i$ and all other representations in the batch, except for the matching pair $(z_i, z_j)$. The objective is for the pair of matching samples to have a higher similarity score compared to the non-matching pairs. For a batch of embedded author records $(z_i, z_j, \ldots, z_n)$, InfoNCE Loss \citep{oord_representation_2018} is defined as:

\begin{equation}\label{eq:contrastive}
\begin{split}
\ell_{i,j} = 
-\log \frac{\exp\left(\operatorname{d}(z_i, z_j)/\tau\right)}
{\sum\limits_{k=1}^{2N} \mathbbm{1}_{[k \neq i]} \exp\left(\operatorname{d}(z_i, z_k)/\tau\right)} \\
= -\frac{\operatorname{d}(z_i, z_j)}{\tau}
+ \log \left[ \sum\limits_{k=1}^{2N} \mathbbm{1}_{[k \neq i]} \exp\left(\frac{\operatorname{d}(z_i, z_k)}{\tau} \right) \right]
\end{split}
\end{equation}

Where $\tau$ is a temperature hyperparameter and $\{z_k\}_{k=1}^{K}$ are embeddings of $K$ negative samples drawn from a noise distribution. The InfoNCE loss encourages the model to embed representations of samples closer together for matching pairs compared to nonmatching pairs within the batch.

Our NAND model employs a feed-forward neural network architecture to learn an embedding space for the disambiguation of pairs. The input varies from an 818 to 3122-dimensional vectors based on the combination of representation methods. The NAND model learns to embed similar samples closer together and dissimilar samples farther apart in the embedding space by minimizing a contrastive loss function. The complete training and inference is outlined in Algorithm~\ref{alg:train}, detailing the steps for embedding author records, loss computation, and similarity classification.

\begin{algorithm}[!t]
\caption{NAND Training and Inference}
\label{alg:train}
\KwIn{Training data $\mathcal{D}_{\text{train}} = \{(r_1^i, r_2^i, y^i)\}_{i=1}^N$, learning rate $\eta$, epochs $T$, threshold $\tau$}
Initialize model parameters $\theta$\;
\textbf{Training Phase}\;
\For{$t = 1, \dots, T$}{
    \For{each $(r_1, r_2, y) \in \mathcal{D}_{\text{train}}$}{
        $\mathbf{e}_1 = \text{Chars2Vec}(r_1) \oplus \text{\textsc{Specter}}(r_1)$\; 
        $\mathbf{e}_2 = \text{Chars2Vec}(r_2) \oplus \text{\textsc{Specter}}(r_2)$\;
        $\mathcal{L} = \text{InfoNCE}(\mathbf{e}_1, \mathbf{e}_2, y)$\; 
        Update $\theta$ using gradient descent with rate $\eta$
    }
}
\textbf{Inference Phase}\;
\For{each $(r_1, r_2) \in \mathcal{D}_{\text{test}}$}{
    $\mathbf{e}_1 = \text{Chars2Vec}(r_1) \oplus \text{\textsc{Specter}}(r_1)$\;
    $\mathbf{e}_2 = \text{Chars2Vec}(r_2) \oplus \text{\textsc{Specter}}(r_2)$\;
    $(\mathbf{o}_1, \mathbf{o}_2) = \text{NAND}_\theta(\mathbf{e}_1, \mathbf{e}_2)$ \;
    $s = \frac{\mathbf{o}_1 \cdot \mathbf{o}_2}{\|\mathbf{o}_1\| \|\mathbf{o}_2\|}$ 
    $\hat{y} = \mathbb{I}(s \ge \tau)$
}
\Return{trained model $\theta$ and predictions $\{\hat{y}\}$}
\end{algorithm}

\begin{table*}[t!]
\centering
\caption{Results of NAND on the LSPO dataset training with Various Representation Methods and Loss Functions. \textbf{Bold} indicates the best results.}
\label{tab:combined_results}
\resizebox{\textwidth}{!}{%
\begin{tabular}{l l c c c c}
\toprule
\textbf{Representation Methods} & \textbf{Loss Functions} & \textbf{Accuracy} & \textbf{Precision} & \textbf{Recall} & \textbf{F1-score} \\
\midrule
Chars2vec + \textsc{Specter} & Cosine-embedding Loss & 91.90\,\% $\pm$ 0.17\,\% & 90.95\,\% $\pm$ 0.24\,\% & \textbf{97.52\,\% $\pm$ 0.07\,\%} & 94.12\,\% $\pm$ 0.12\,\% \\
Chars2vec + \textsc{Specter} & InfoNCE Loss & \textbf{94.64\,\% $\pm$ 0.04\,\%} & \textbf{96.67\,\% $\pm$ 0.05\,\%} & 95.21\,\% $\pm$ 0.11\,\% & \textbf{95.94\,\% $\pm$ 0.03\,\%} \\
Chars2vec + Llama 3.2 & Cosine-embedding Loss & 92.12\,\% $\pm$ 0.09\,\% & 91.48\,\% $\pm$ 0.19\,\% & 97.21\,\% $\pm$ 0.12\,\% & 94.25\,\% $\pm$ 0.05\,\% \\
Chars2vec + Llama 3.2 & InfoNCE Loss & 93.66\,\% $\pm$ 0.09\,\% & 95.57\,\% $\pm$ 0.12\,\% & 94.86\,\% $\pm$ 0.18\,\% & 95.22\,\% $\pm$ 0.07\,\% \\
Word2vec + \textsc{Specter} & Cosine-embedding Loss & 86.89\,\% $\pm$ 0.07\,\% & 88.72\,\% $\pm$ 0.60\,\% & 92.01\,\% $\pm$ 0.69\,\% & 90.31\,\% $\pm$ 0.06\,\% \\
Word2vec + \textsc{Specter} & InfoNCE Loss & 90.21\,\% $\pm$ 0.13\,\% & 92.01\,\% $\pm$ 0.25\,\% & 93.39\,\% $\pm$ 0.09\,\% & 92.69\,\% $\pm$ 0.08\,\% \\
Word2vec + Llama 3.2 & Cosine-embedding Loss & 90.27\,\% $\pm$ 0.18\,\% & 89.58\,\% $\pm$ 0.20\,\% & 96.59\,\% $\pm$ 0.09\,\% & 92.95\,\% $\pm$ 0.12\,\% \\
Word2vec + Llama 3.2 & InfoNCE Loss & 92.64\,\% $\pm$ 0.04\,\% & 95.63\,\% $\pm$ 0.06\,\% & 93.20\,\% $\pm$ 0.06\,\% & 94.39\,\% $\pm$ 0.03\,\% \\
\bottomrule
\end{tabular}%
}
\end{table*}

\subsection{Clustering}\label{subsec:clustering}
After training, the NAND model produces embeddings for each author record. We then cluster these embeddings to partition records into unique researcher identities in the learned latent space. Let \(\mathcal{Z} = \{z_1, z_2, \dots, z_n\}\) denote the set of learned embeddings for all author records in a specific name block \(B_{k}\). The objective of clustering is to group embeddings in \(Z_{k}\) such that each cluster corresponds to a unique researcher identity. Clustering within name blocks allows parallelization and avoids comparing unrelated names (e.g., J. Smith vs. Y. Wang), which is less meaningful. 

AND requires an algorithm that does not assume the number of clusters a priori, since the number of unique researchers is unknown. In line with prior work \citep{khabsa_large_2014}, DBSCAN performs best in our learned embedding space. DBSCAN groups points based on local density in the embedding space, using two hyperparameters: the neighborhood radius $\epsilon$ and the minimum number of points $m$ required to form a dense region \citep{ester_density-based_1996}. Records within an $\epsilon$-neighborhood of at least $m$ points are assigned to the same cluster, while sparse points are treated as noise. We compute distances using cosine distance, consistent with the metric used during NAND training and evaluation.

In line with prior AND work \citep{khabsa_online_2015}, we introduce task-specific constraints to improve clustering reliability. When full name metadata is available (e.g., John J. Smith vs. John H. Smith), records with conflicting full names cannot be merged, even if their embeddings are close. Additionally, clusters are constrained to avoid grouping records separated by implausibly long publication gaps (e.g., \num{30} years), reducing errors caused by embedding similarity.

\section{Experimental Setup} \label{sec:experiment}
In this section, we describe training setup and evaluation metrics used to assess the performance of the NAND algorithm. We present the architecture for the similarity learning model and the hyperparameters for the clustering algorithm.

We trained and evaluated the NAND model using a systematic approach designed to ensure reproducibility and robust performance evaluation. We performed hyperparameter optimization separately for each representation method using the Optuna\footnote{\url{https://optuna.org/}} Python package, with 100–150 trials per configuration. We employed a grid sampler to explore the search space, targeting architectural and training parameters such as hidden layer width, number of layers (depth), activation functions, learning rate, dropout rate, batch size, and size of output embedding (see Table~\ref{tab:hyperparams}).

The base model is a feed-forward neural network. Input embeddings (i.e., from the respective representation methods) are passed through fully connected layers, each followed by non-linear activations (e.g., LeakyReLU or SELU, depending on the tuned configuration), and dropout layers to reduce overfitting. The final hidden layer produces low-dimensional representations, which are compared using cosine similarity for the pairwise training objective. We set a cosine similarity threshold to decide whether a pair represents the same author. The threshold is chosen by analyzing the Receiver Operating Characteristic (ROC) curve on the validation set, ensuring the best balance between precision and recall.

Utilizing the optimized hyperparameters, we trained each representation method five times until convergence, with different random seeds to account for variations from data shuffling and weight initialization. During training, we employed early stopping with a patience of \num{25} epochs, monitoring validation loss to prevent overfitting and saving the best-performing epoch from each run. To evaluate model performance, we tested the best epoch from each of the five runs on the held-out test set. We computed multiple performance metrics including accuracy, recall, precision, and F1 score, reporting their means and standard errors across the runs.

For clustering, hyperparameters were tuned empirically for each block of papers. Larger blocks required smaller $\epsilon$ values, while smaller blocks performed better with larger values, with $\epsilon$ ranging from \num{0.2} to \num{0.5},  consistent with the density variations across blocks. We fixed the minimum number of samples parameter to $m=1$ to account for researchers who only contribute to a single publication. In our experiments, clustering performance is more sensitive to the choice of $\epsilon$ than to $m$.

\section{Results} \label{sec:results} 
We evaluate the NAND pairwise similarity model on the LSPO dataset, followed by clustering, ablation studies, and cross-domain experiments.  

\subsection{Pairwise Similarity on LSPO}  
We report the evaluation of various author record representation methods and contrastive loss functions for the training of the NAND model on the LSPO dataset in Table~\ref{tab:combined_results}. The combination of chars2vec, \textsc{Specter}, and InfoNCE loss achieved the highest overall performance, with an accuracy of 94.64\%, precision of 95.21\%, recall of 96.67\%, and an F1-score of 95.94\% on a balanced test set of hard negatives and positives, considering only pairs sampled from within the same name block. However, it is noteworthy that the combination of chars2vec, \textsc{Specter}, and cosine-embedding loss attained a higher recall of 97.52\%. In terms of representing researcher information, the chars2vec representation consistently performed higher than Word2vec across all transformer models and loss functions. The highest-performing Word2vec model was in combination with Llama 3.2 and InfoNCE loss, achieving an accuracy of 92.64\%, precision of 95.63\%, recall of 93.20\%, and F1-score of 94.39\%. These results suggest that character-level embeddings are more suitable for the AND task on the LSPO dataset, potentially due to their ability to better handle variations, abbreviations, misspellings, or parsing errors in author names across a large digital library. Concerning the representation of publication content (i.e., abstracts and titles), the Llama 3.2 model outperforms \textsc{Specter} across all similar experiments, except the highest-performing model. Notably, the \textsc{Specter} model is trained with the objective of generating document-level embeddings of scientific documents, ensuring that publications with similar content are represented more similarly, which translates to our AND task formulated as a similarity learning problem. However, Llama 3.2 does perform well across various metrics.

\begin{table*}[t!]
\centering
\centering
\caption{Clustering performance of NAND embeddings using DBSCAN on the LSPO dataset.}
\label{tab:clustering_results}
\resizebox{\linewidth}{!}{%
\begin{tabular}{l c c c c}
\toprule
\textbf{Metric} & \textbf{Accuracy} & \textbf{Precision} & \textbf{Recall} & \textbf{F1-score} \\
\midrule
DBSCAN & {88.42 \,\% $\pm$ 0.03\,\%} & {95.75 \,\% $\pm$ 0.01\,\%} & \textbf{{96.67 \,\% $\pm$ 0.01\,\%}} & {95.88 \,\% $\pm$ 0.00\,\%} \\
DBSCAN + constraints & \textbf{{91.91 \,\% $\pm$ 0.02\,\%}} & \textbf{{96.15 \,\% $\pm$ 0.01\,\%}} & {96.21 \,\% $\pm$ 0.01\,\%} & \textbf{{95.93 \,\% $\pm$ 0.01,\%}} \\
\bottomrule
\end{tabular}%
}
\end{table*}

\subsection{Clustering Performance on LSPO}  
We evaluated the clustering performance of the trained NAND embeddings on the LSPO dataset using the DBSCAN with constraints algorithm described in Subsection~\ref{subsec:clustering}. Clustering was performed within name blocks and assessed with pairwise clustering metrics following prior work \citep{zhang_name_2018}. Without constraints, DBSCAN achieved a cluster-level accuracy (hereafter referred to as clustering accuracy) of 88.42\%, defined as the percentage of clusters that perfectly matched the ground truth. Pairwise evaluation showed precision of 95.75\%, recall of 96.67\%, and an F1-score of 95.88\%. Incorporating domain-specific constraints, full-name consistency and temporal ordering, further improved performance, raising accuracy to 91.91\% and F1-score to 95.93\%, while maintaining high precision of 96.15\% and recall of 96.21\%. Singleton clusters were included in evaluation, ensuring that researchers with only one publication were correctly represented. These results show that learned NAND embeddings can be effectively clustered into disjoint researcher identities, and that combining DBSCAN with domain-specific constraints provides a robust and scalable solution that balances both precision and recall. 

\subsection{Ablation}\label{subsec:ablation} 
We examined the roles of the underlying architecture and representation techniques on performance. The architecture and representation techniques, even without training, contribute significantly to the NAND's performance. As shown in Table~\ref{tab:notrain}, when we evaluated the model without any training for pairwise similarity, using only the initial embeddings and architecture, we found that it achieved approximately 85\% accuracy. This suggests that the choice of representation technique is crucial for the AND task. While training improves the accuracy by approximately 10\%, the strong baseline performance indicates that effectively representing author data through appropriate embedding techniques is fundamental to distinguishing between author identities.

Furthermore, we conducted experiments incorporating affiliation and venue information into NAND training to better understand the contributions of additional metadata sources. We encoded affiliations with a chars2vec model to capture institutional variations and abbreviations (e.g., “Michigan State University” vs. “MSU”). Publishing patterns were captured by applying \textsc{Specter} to publication venue/journal names generating semantic embeddings that reflect relatedness across subfields. As shown in Table~\ref{tab:ablation_pairwise}, for pairwise similarity, adding affiliation metadata slightly reduced accuracy to 93.65\% and recall to 93.53\%, but achieved the highest precision of 96.81\%. Incorporating venue metadata produced a more balanced performance, with accuracy reaching 94.41\%, recall 95.59\%, precision 95.99\%, and an F1-score of 95.79\%. For clustering (Table~\ref{tab:ablation_clustering}), affiliations slightly lowered accuracy to 91.16\% and F1-score to 95.80\%, while maintaining strong precision at 96.05\%. Venue metadata decreased accuracy to 90.43\%, but yielded the highest recall of 96.25\%.

Motivated by prior work that leverages author name, title, and publication venue as consistently available metadata \citep{ferreira_disambiguating_2014}, we evaluated the impact of removing abstracts during training and evaluation. Using title-only \textsc{Specter} embeddings led to a modest reduction in performance. Pairwise similarity accuracy decreased from 94.41\% to 93.73\% without abstracts, and clustering accuracy dropped from 90.43\% to 90.15\%.  While publication venue data is generally more available than abstracts in large-scale digital libraries, incomplete metadata can be augmented through sources such as OpenAlex. These results show that although abstracts provide measurable benefit, our approach remains robust across various metadata conditions. Overall, the ablation experiments highlight the challenges of relying on sparse, noisy, or heterogeneous metadata at scale.

\begin{table*}[t!]
\centering
\caption{Ablation results on pairwise similarity using the LSPO dataset. Each row corresponds to a separate experiment with different metadata features, indicated in the “Metadata” column.}
\label{tab:ablation_pairwise}
\resizebox{\textwidth}{!}{%
\begin{tabular}{l l c c c}
\toprule
\textbf{Metadata} & \textbf{Accuracy} & \textbf{Precision} & \textbf{Recall} & \textbf{F1-score} \\
\midrule
Baseline (Title and Abstract) & \textbf{94.64\,\% $\pm$ 0.04\,\%} & 96.67\,\% $\pm$ 0.05\,\% & 95.21\,\% $\pm$ 0.11\,\% & \textbf{95.94\,\% $\pm$ 0.03\,\%} \\
+ Affiliation & 93.65\,\% $\pm$ 0.05\,\% & \textbf{96.81\,\% $\pm$ 0.11\,\%} & 93.53\,\% $\pm$ 0.05\,\% & 95.14\,\% $\pm$ 0.04\,\% \\
+ Publication Venue  & 94.41\,\% $\pm$ 0.04\,\% & 95.99\,\% $\pm$ 0.14\,\% & \textbf{95.59\,\% $\pm$ 0.20\,\%} & 95.79\,\% $\pm$ 0.03\,\% \\
+ Publication Venue - Abstract & 93.73\,\% $\pm$ 0.09\,\% & 95.29\,\% $\pm$ 0.17\,\% & 95.29\,\% $\pm$ 0.32\,\% & 95.28\,\% $\pm$ 0.08\,\% \\
\bottomrule
\end{tabular}%
}
\end{table*}

\subsection{Cross-Domain Evaluation on LAGOS-AND}
We trained and evaluated the NAND framework on the cross-domain LAGOS-AND dataset to assess whether the results extend beyond the physics and astronomy literature. We observe in Table~\ref{tab:train_test_LAGOS-AND} that both models (i.e., training on LSPO and LAGOS-AND) achieve strong performance (i.e., >95\% F1-Score) when evaluated on their respective training datasets. However, when each model is evaluated on the opposite dataset (e.g., NAND trained on LSPO and evaluated on LAGOS-AND and vice versa), we observe a noticeable decrease in performance of between 3-4\%. This performance gap highlights two key findings: first, that domain-specific training yields optimal results for field-specific disambiguation tasks, and second, that there is a clear need to maintain both specialized and general-domain AND data to effectively serve different scholarly tasks.

\section{Discussion}\label{sec:discussion} 
We focus on comparing NAND to approaches that are applicable to real-world large digital libraries. Specifically, we compare to approaches which do not rely on extensive metadata that is missing or erroneous. 
We first compare against the hybrid-initials method of \citet{milojevic_accuracy_2013}, which uses surname frequency to choose between first-only or middle-initial name matching. While effective for small name blocks, it produces merging and splitting errors for common names. On the LSPO dataset, it over-merges large name blocks and achieves only 75.24
\% accuracy across over \num{125000} ORCiD-linked identities. In contrast, NAND reaches 91.91\% clustering accuracy on the same dataset. 
We also compare against a random forest classification and clustering approach that was trained on a manually labeled NASA/ADS subset \citep{mihaljevic_disambiguation_2021}. The random forest method achieves a clustering F1 score of 94.7\% on \num{9545} records from \num{562} researchers across \num{39} name blocks. NAND achieves 95.93\% F1 on the much larger LSPO dataset, also derived from NASA/ADS.
Beyond NASA/ADS, we consider algorithms designed for curated benchmarks such as Aminer-18, which covers the 100 most ambiguous Chinese names with full metadata (e.g., affiliations, keywords). Methods tailored to this dataset perform well on the curated dataset but as they rely on extensive metadata we can not test their performance on our LSPO datset. We thus apply our method on the Aminer-18 dataset for comparison. The first method \citep{liu_author_2024} reports 78.09\% precision and 82.32\% recall. A second method that utilizes only coauthor graphs \citep{fan_graph-based_2011} achieves 81.62\% precision and 40.43\% recall. NAND yields 60.83\% precision and 66.23\% recall on the Aminer-18 dataset. These results highlight a key trade-off: on the one hand, NAND does not match algorithms that exploit full metadata on curated datasets, on the other such methods do not work on real-world datasets. Furthermore, multiple studies confirm that strong benchmark results do not necessarily translate to real-world performance \citep{subramanian_s2and_2021, kim_model_2020}. We note, that missing affiliation data systematically degrades large-scale disambiguation, reinforcing that benchmarks with full coverage overestimate performance \citep{zhou_evaluating_2024}. We therefore present Aminer-18 comparisons for completeness, but emphasize that NAND is intended for realistic settings where metadata is sparse or inconsistent and is tested against such datasets. 

\section{Conclusion}\label{sec:conclusion}
In conclusion, this work presents the Neural Author Name Disambiguator (NAND), a scalable and practical solution for accurately linking research papers to their respective authors across large digital libraries. A key contribution of NAND is its minimal reliance on metadata making it applicable to large digital libraries where detailed metadata required by current AND solutions is incomplete. We also share the Large-Scale ORCiD-Linked Physics Dataset (LSPO) for Author Name Disambiguation to develop and evaluate NAND. LSPO links \num{553496} publications from NASA/ADS to \num{125486} ORCID profiles, addressing the size and coverage limitations of existing physics and astronomy datasets. NAND embeds researcher and publication metadata into a latent space by learning to represent records from the same researcher closer together, while pushing apart records from different researchers. NAND achieves 94.64\% accuracy on the LSPO dataset. Our results demonstrate that high-accuracy disambiguation is possible with minimal metadata, delivering a scalable, practical solution to the widespread problem of name ambiguity in real-world digital libraries such as NASA/ADS.

\begin{table*}[t!]
\centering
\caption{Ablation results on clustering using the LSPO dataset. Each row corresponds to a separate experiment with different metadata features, indicated in the “Metadata” column.}
\label{tab:ablation_clustering}
\resizebox{\textwidth}{!}{%
\begin{tabular}{l l c c c }
\toprule
\textbf{Metadata} & \textbf{Accuracy} & \textbf{Precision} & \textbf{Recall} & \textbf{F1-score} \\
\midrule
Baseline (Title and Abstract) & \textbf{{91.91 \,\% $\pm$ 0.02\,\%}} & \textbf{{96.15 \,\% $\pm$ 0.00\,\%}} & {96.21 \,\% $\pm$ 0.01\,\%} & \textbf{{95.93 \,\% $\pm$ 0.01,\%}} \\
+ Affiliation & 91.16\,\% $\pm$ 0.08\,\% & 96.05\,\% $\pm$ 0.02\,\% & 96.09\,\% $\pm$ 0.03\,\% & 95.80\,\% $\pm$ 0.03\,\% \\
+ Publication Venue  & 90.43\,\% $\pm$ 0.04\,\% & 95.97\,\% $\pm$ 0.01\,\% & \textbf{96.25\,\% $\pm$ 0.01\,\%} & 95.82\,\% $\pm$ 0.01\,\% \\
+ Publication Venue - Abstract & 90.15\,\% $\pm$ 0.05\,\% & 95.83\,\% $\pm$ 0.01\,\% & 96.13\,\% $\pm$ 0.04\,\% & 95.68\,\% $\pm$ 0.01\,\% \\
\bottomrule
\end{tabular}%
}
\end{table*}

\begin{table*}[t!]
\centering
\caption{F1-Score Performance For Different Training Sets, Evaluated On Different Target Test Sets. \textbf{Bold} indicates the best results.}
\label{tab:train_test_LAGOS-AND}
\begin{tabular}{l c c} 
\toprule
Train $\downarrow$ / Test $\rightarrow$ & LSPO & LAGOS-AND \\
\midrule
LSPO & \textbf{95.94\,\% $\pm$ 0.03\,\%} & 93.36\,\% $\pm$ 0.01\,\% \\
LAGOS-AND & 92.61\,\% $\pm$ 0.12\,\% & \textbf{97.69\,\% $\pm$ 0.02\,\%} \\
\bottomrule
\end{tabular}
\end{table*}

\paragraph{\textbf{Acknowledgments}} We acknowledge the use of data from the SAO/NASA Astrophysics Data System API and the ORCiD public data file, which provided open access to the research resources essential for this work. This research was supported in part by computational resources and services from the Institute for Cyber-Enabled Research at Michigan State University. 
Additionally, this work was supported by the National Science Foundation Research Traineeship Program (DGE-2152014) for Vicente Amado Olivo. We also gratefully acknowledge funding support from the European Space Agency (ESA) through a traineeship for Vicente Amado Olivo, along with the guidance and support of Markus Kissler-Patig and Jan Reerink at ESA.

\paragraph{\textbf{Contributor Roles}}
\begin{enumerate}
    \item Conceptualization: Vicente Amado Olivo, Nutan Chen, Wolfgang Kerzendorf
    \item Data Curation: Vicente Amado Olivo, Joshua V. Shields, Bangjing Lu, Andreas Flors
    \item Formal Analysis: Vicente Amado Olivo
    \item Funding Acquisition: Wolfgang Kerzendorf
    \item Investigation: Vicente Amado Olivo, Nutan Chen
    \item Methodology: Vicente Amado Olivo,  Nutan Chen, Wolfgang Kerzendorf
    \item Project Administration: Wolfgang Kerzendorf
    \item Resources: Wolfgang Kerzendorf
    \item Software: Vicente Amado Olivo
    \item Supervision: Wolfgang Kerzendorf
    \item Validation: Vicente Amado Olivo
    \item Visualization: Vicente Amado Olivo
    \item Writing - original draft: Vicente Amado Olivo
    \item Writing - reviewing \& editing: Vicente Amado Olivo, Wolfgang Kerzendorf, Nutan Chen, Josh Shields
\end{enumerate}

\bibliography{Neural_AND}

\begin{thebibliography}{}
\expandafter\ifx\csname natexlab\endcsname\relax\def\natexlab#1{#1}\fi
\providecommand{\url}[1]{\href{#1}{#1}}
\providecommand{\dodoi}[1]{doi:~\href{http://doi.org/#1}{\nolinkurl{#1}}}
\providecommand{\doeprint}[1]{\href{http://ascl.net/#1}{\nolinkurl{http://ascl.net/#1}}}
\providecommand{\doarXiv}[1]{\href{https://arxiv.org/abs/#1}{\nolinkurl{https://arxiv.org/abs/#1}}}

\bibitem[{{A. J. Lotka}(1926)}]{a_j_lotka_frequency_1926}
{A. J. Lotka}. 1926, Journal of the Franklin Institute, 202, 271, \dodoi{10.1016/S0016-0032(26)91166-6}

\bibitem[{Bello \& Galindo-Rueda(2020)}]{bello_charting_2020}
Bello, M., \& Galindo-Rueda, F. 2020, Charting the digital transformation of science: {Findings} from the 2018 {OECD} {International} {Survey} of {Scientific} {Authors} ({ISSA2}), Tech. rep., OECD, Paris, \dodoi{10.1787/1b06c47c-en}

\bibitem[{Borgman \& Siegfried(1992)}]{borgman_gettys_nodate}
Borgman, C.~L., \& Siegfried, S.~L. 1992, Getty's {Synoname}™ and its cousins: {A} survey of applications of personal name‐matching algorithms - {Borgman} - 1992 - {Journal} of the {American} {Society} for {Information} {Science} - {Wiley} {Online} {Library}, \dodoi{10.1002/(SICI)1097-4571(199208)43:7<459::AID-ASI1>3.0.CO;2-D}

\bibitem[{Boukhers \& Asundi(2023)}]{boukhers_deep_2023}
Boukhers, Z., \& Asundi, N.~B. 2023, International Journal on Digital Libraries, \dodoi{10.1007/s00799-023-00361-6}

\bibitem[{Breiman(2001)}]{breiman_random_2001}
Breiman, L. 2001, Machine Learning, 45, 5, \dodoi{10.1023/A:1010933404324}

\bibitem[{Cao \& Rei(2016)}]{cao_joint_2016}
Cao, K., \& Rei, M. 2016, in Proceedings of the 1st {Workshop} on {Representation} {Learning} for {NLP} (Berlin, Germany: Association for Computational Linguistics), 18--26, \dodoi{10.18653/v1/W16-1603}

\bibitem[{Chen {et~al.}(2023)Chen, Zhang, Zhang, Han, Cheng, Li, Dong, \& Tang}]{chen_web-scale_2023}
Chen, B., Zhang, J., Zhang, F., {et~al.} 2023, in Proceedings of the 29th {ACM} {SIGKDD} {Conference} on {Knowledge} {Discovery} and {Data} {Mining} (Long Beach CA USA: ACM), 3817--3828, \dodoi{10.1145/3580305.3599930}

\bibitem[{Choudhury {et~al.}(2023)Choudhury, Salsabil, Jayanetti, Wu, Ingram, \& Fox}]{choudhury_metaenhance_2023}
Choudhury, M.~H., Salsabil, L., Jayanetti, H.~R., {et~al.} 2023, {MetaEnhance}: {Metadata} {Quality} {Improvement} for {Electronic} {Theses} and {Dissertations} of {University} {Libraries} | {IEEE} {Conference} {Publication} | {IEEE} {Xplore}.
\newblock \url{https://ieeexplore.ieee.org/document/10265916}

\bibitem[{Cohan {et~al.}(2020)Cohan, Feldman, Beltagy, Downey, \& Weld}]{cohan_specter_2020}
Cohan, A., Feldman, S., Beltagy, I., Downey, D., \& Weld, D. 2020, Proceedings of the 58th Annual Meeting of the Association for Computational Linguistics, 2270, \dodoi{10.18653/v1/2020.acl-main.207}

\bibitem[{D'Angelo {et~al.}(2011)D'Angelo, Giuffrida, \& Abramo}]{dangelo_heuristic_2011}
D'Angelo, C.~A., Giuffrida, C., \& Abramo, G. 2011, Journal of the American Society for Information Science and Technology, 62, 257, \dodoi{10.1002/asi.21460}

\bibitem[{Eggert(2011)}]{eggert_best_2011}
Eggert, L.~D. 2011, Frontiers in Psychology, 2, 196, \dodoi{10.3389/fpsyg.2011.00196}

\bibitem[{Ester {et~al.}(1996)Ester, Kriegel, Sander, \& Xu}]{ester_density-based_1996}
Ester, M., Kriegel, H., Sander, J., \& Xu, X. 1996, in A {Density}-{Based} {Algorithm} for {Discovering} {Clusters} in {Large} {Spatial} {Databases} with {Noise}.
\newblock \url{https://dl.acm.org/doi/10.5555/3001460.3001507}

\bibitem[{Fan {et~al.}(2011)Fan, Wang, Pu, Zhou, \& Lv}]{fan_graph-based_2011}
Fan, X., Wang, J., Pu, X., Zhou, L., \& Lv, B. 2011, Journal of Data and Information Quality, 2, 1, \dodoi{10.1145/1891879.1891883}

\bibitem[{Ferreira {et~al.}(2014)Ferreira, Gonçalves, \& Laender}]{ferreira_disambiguating_2014}
Ferreira, A.~A., Gonçalves, M.~A., \& Laender, A. H.~F. 2014, World Digital Libraries, 7, 71, \dodoi{10.3233/WDL-120115}

\bibitem[{Ferreira {et~al.}(2010)Ferreira, Veloso, Gonçalves, \& Laender}]{ferreira_effective_2010}
Ferreira, A.~A., Veloso, A., Gonçalves, M.~A., \& Laender, A.~H. 2010, in Proceedings of the 10th annual joint conference on {Digital} libraries (Gold Coast Queensland Australia: ACM), 39--48, \dodoi{10.1145/1816123.1816130}

\bibitem[{Grattafiori {et~al.}(2024)Grattafiori, Dubey, Jauhri, Pandey, Kadian, Al-Dahle, Letman, Mathur, Schelten, Vaughan, Yang, Fan, Goyal, Hartshorn, Yang, Mitra, Sravankumar, Korenev, Hinsvark, Rao, Zhang, Rodriguez, Gregerson, Spataru, Roziere, Biron, Tang, Chern, Caucheteux, Nayak, Bi, Marra, McConnell, Keller, Touret, Wu, Wong, Ferrer, Nikolaidis, Allonsius, Song, Pintz, Livshits, Wyatt, Esiobu, Choudhary, Mahajan, Garcia-Olano, Perino, Hupkes, Lakomkin, AlBadawy, Lobanova, Dinan, Smith, Radenovic, Guzmán, Zhang, Synnaeve, Lee, Anderson, Thattai, Nail, Mialon, Pang, Cucurell, Nguyen, Korevaar, Xu, Touvron, Zarov, Ibarra, Kloumann, Misra, Evtimov, Zhang, Copet, Lee, Geffert, Vranes, Park, Mahadeokar, Shah, van~der Linde, Billock, Hong, Lee, Fu, Chi, Huang, Liu, Wang, Yu, Bitton, Spisak, Park, Rocca, Johnstun, Saxe, Jia, Alwala, Prasad, Upasani, Plawiak, Li, Heafield, Stone, El-Arini, Iyer, Malik, Chiu, Bhalla, Lakhotia, Rantala-Yeary, van~der Maaten, Chen, Tan, Jenkins, Martin, Madaan, Malo, Blecher,
  Landzaat, de~Oliveira, Muzzi, Pasupuleti, Singh, Paluri, Kardas, Tsimpoukelli, Oldham, Rita, Pavlova, Kambadur, Lewis, Si, Singh, Hassan, Goyal, Torabi, Bashlykov, Bogoychev, Chatterji, Zhang, Duchenne, Çelebi, Alrassy, Zhang, Li, Vasic, Weng, Bhargava, Dubal, Krishnan, Koura, Xu, He, Dong, Srinivasan, Ganapathy, Calderer, Cabral, Stojnic, Raileanu, Maheswari, Girdhar, Patel, Sauvestre, Polidoro, Sumbaly, Taylor, Silva, Hou, Wang, Hosseini, Chennabasappa, Singh, Bell, Kim, Edunov, Nie, Narang, Raparthy, Shen, Wan, Bhosale, Zhang, Vandenhende, Batra, Whitman, Sootla, Collot, Gururangan, Borodinsky, Herman, Fowler, Sheasha, Georgiou, Scialom, Speckbacher, Mihaylov, Xiao, Karn, Goswami, Gupta, Ramanathan, Kerkez, Gonguet, Do, Vogeti, Albiero, Petrovic, Chu, Xiong, Fu, Meers, Martinet, Wang, Wang, Tan, Xia, Xie, Jia, Wang, Goldschlag, Gaur, Babaei, Wen, Song, Zhang, Li, Mao, Coudert, Yan, Chen, Papakipos, Singh, Srivastava, Jain, Kelsey, Shajnfeld, Gangidi, Victoria, Goldstand, Menon, Sharma, Boesenberg,
  Baevski, Feinstein, Kallet, Sangani, Teo, Yunus, Lupu, Alvarado, Caples, Gu, Ho, Poulton, Ryan, Ramchandani, Dong, Franco, Goyal, Saraf, Chowdhury, Gabriel, Bharambe, Eisenman, Yazdan, James, Maurer, Leonhardi, Huang, Loyd, De~Paola, Paranjape, Liu, Wu, Ni, Hancock, Wasti, Spence, Stojkovic, Gamido, Montalvo, Parker, Burton, Mejia, Liu, Wang, Kim, Zhou, Hu, Chu, Cai, Tindal, Feichtenhofer, Gao, Civin, Beaty, Kreymer, Li, Adkins, Xu, Testuggine, David, Parikh, Liskovich, Foss, Wang, Le, Holland, Dowling, Jamil, Montgomery, Presani, Hahn, Wood, Le, Brinkman, Arcaute, Dunbar, Smothers, Sun, Kreuk, Tian, Kokkinos, Ozgenel, Caggioni, Kanayet, Seide, Florez, Schwarz, Badeer, Swee, Halpern, Herman, Sizov, Guangyi, Zhang, Lakshminarayanan, Inan, Shojanazeri, Zou, Wang, Zha, Habeeb, Rudolph, Suk, Aspegren, Goldman, Zhan, Damlaj, Molybog, Tufanov, Leontiadis, Veliche, Gat, Weissman, Geboski, Kohli, Lam, Asher, Gaya, Marcus, Tang, Chan, Zhen, Reizenstein, Teboul, Zhong, Jin, Yang, Cummings, Carvill, Shepard, McPhie,
  Torres, Ginsburg, Wang, Wu, U, Saxena, Khandelwal, Zand, Matosich, Veeraraghavan, Michelena, Li, Jagadeesh, Huang, Chawla, Huang, Chen, Garg, A, Silva, Bell, Zhang, Guo, Yu, Moshkovich, Wehrstedt, Khabsa, Avalani, Bhatt, Mankus, Hasson, Lennie, Reso, Groshev, Naumov, Lathi, Keneally, Liu, Seltzer, Valko, Restrepo, Patel, Vyatskov, Samvelyan, Clark, Macey, Wang, Hermoso, Metanat, Rastegari, Bansal, Santhanam, Parks, White, Bawa, Singhal, Egebo, Usunier, Mehta, Laptev, Dong, Cheng, Chernoguz, Hart, Salpekar, Kalinli, Kent, Parekh, Saab, Balaji, Rittner, Bontrager, Roux, Dollar, Zvyagina, Ratanchandani, Yuvraj, Liang, Alao, Rodriguez, Ayub, Murthy, Nayani, Mitra, Parthasarathy, Li, Hogan, Battey, Wang, Howes, Rinott, Mehta, Siby, Bondu, Datta, Chugh, Hunt, Dhillon, Sidorov, Pan, Mahajan, Verma, Yamamoto, Ramaswamy, Lindsay, Feng, Lin, Zha, Patil, Shankar, Zhang, Wang, Agarwal, Sajuyigbe, Chintala, Max, Chen, Kehoe, Satterfield, Govindaprasad, Gupta, Deng, Cho, Virk, Subramanian, Choudhury, Goldman, Remez,
  Glaser, Best, Koehler, Robinson, Li, Zhang, Matthews, Chou, Shaked, Vontimitta, Ajayi, Montanez, Mohan, Kumar, Mangla, Ionescu, Poenaru, Mihailescu, Ivanov, Li, Wang, Jiang, Bouaziz, Constable, Tang, Wu, Wang, Wu, Gao, Kleinman, Chen, Hu, Jia, Qi, Li, Zhang, Zhang, Adi, Nam, Yu, {Wang}, Zhao, Hao, Qian, Li, He, Rait, DeVito, Rosnbrick, Wen, Yang, Zhao, \& Ma}]{grattafiori_llama_2024}
Grattafiori, A., Dubey, A., Jauhri, A., {et~al.} 2024, \dodoi{10.48550/ARXIV.2407.21783}

\bibitem[{Gurney {et~al.}(2011)Gurney, Horlings, \& van~den Besselaar}]{gurney_author_2011}
Gurney, T., Horlings, E., \& van~den Besselaar, P. 2011, Scientometrics, 91, 435, \dodoi{10.1007/s11192-011-0589-1}

\bibitem[{Han {et~al.}(2004)Han, Giles, Zha, Li, \& Tsioutsiouliklis}]{han_two_2004}
Han, H., Giles, L., Zha, H., Li, C., \& Tsioutsiouliklis, K. 2004, in Proceedings of the 4th {ACM}/{IEEE}-{CS} joint conference on {Digital} libraries (Tuscon AZ USA: ACM), 296--305, \dodoi{10.1145/996350.996419}

\bibitem[{Imambi {et~al.}(2021)Imambi, Prakash, \& Kanagachidambaresan}]{imambi2021pytorch}
Imambi, S., Prakash, K.~B., \& Kanagachidambaresan, G. 2021, Programming with TensorFlow: Solution for Edge Computing Applications, 87

\bibitem[{Ioannidis {et~al.}(2014)Ioannidis, Boyack, \& Klavans}]{noauthor_estimates_nodate}
Ioannidis, J.~P., Boyack, K.~W., \& Klavans, R. 2014, PloS one, 9, e101698

\bibitem[{Kerzendorf(2019)}]{kerzendorf_knowledge_2019}
Kerzendorf, W.~E. 2019, Journal of Astrophysics and Astronomy, 40, 23, \dodoi{10.1007/s12036-019-9590-5}

\bibitem[{Khabsa {et~al.}(2014)Khabsa, Treeratpituk, \& Giles}]{khabsa_large_2014}
Khabsa, M., Treeratpituk, P., \& Giles, C.~L. 2014, in 2014 {IEEE} {International} {Conference} on {Big} {Data} ({Big} {Data}) (Washington, DC, USA: IEEE), 41--42, \dodoi{10.1109/BigData.2014.7004487}

\bibitem[{Khabsa {et~al.}(2015)Khabsa, Treeratpituk, \& Giles}]{khabsa_online_2015}
Khabsa, M., Treeratpituk, P., \& Giles, C.~L. 2015, Proceedings of the 15th ACM/IEEE-CS Joint Conference on Digital Libraries, 37, \dodoi{10.1145/2756406.2756915}

\bibitem[{Kim(2018)}]{kim_evaluating_2018}
Kim, J. 2018, Scientometrics, 116, 1867, \dodoi{10.1007/s11192-018-2824-5}

\bibitem[{Kim \& Owen-Smith(2020)}]{kim_model_2020}
Kim, J., \& Owen-Smith, J. 2020, IEEE Access, 8, 188378, \dodoi{10.1109/ACCESS.2020.3031112}

\bibitem[{Kim \& Owen-Smith(2021)}]{kim_orcid-linked_2021}
---. 2021, Scientometrics, 126, 2057, \dodoi{10.1007/s11192-020-03826-6}

\bibitem[{Koh {et~al.}(2021)Koh, Sagawa, Marklund, Xie, Zhang, Balsubramani, Hu, Yasunaga, Phillips, Gao, Lee, David, Stavness, Guo, Earnshaw, Haque, Beery, Leskovec, Kundaje, Pierson, Levine, Finn, \& Liang}]{koh_wilds_2021}
Koh, P.~W., Sagawa, S., Marklund, H., {et~al.} 2021, {WILDS}: {A} {Benchmark} of in-the-{Wild} {Distribution} {Shifts},  arXiv, \dodoi{10.48550/arXiv.2012.07421}

\bibitem[{Kurtz {et~al.}(2005)Kurtz, Eichhorn, Accomazzi, Grant, Demleitner, Henneken, \& Murray}]{kurtz_effect_2005}
Kurtz, M.~J., Eichhorn, G., Accomazzi, A., {et~al.} 2005, Information Processing \& Management, 41, 1395, \dodoi{10.1016/j.ipm.2005.03.010}

\bibitem[{Liu {et~al.}(2024)Liu, Zhang, Chen, \& Chen}]{liu_author_2024}
Liu, D., Zhang, R., Chen, J., \& Chen, X. 2024, in Proceedings of the {ACM} {Web} {Conference} 2024, {WWW} '24 (New York, NY, USA: Association for Computing Machinery), 2193--2203, \dodoi{10.1145/3589334.3645596}

\bibitem[{Louppe {et~al.}(2016)Louppe, Al-Natsheh, Susik, \& Maguire}]{ngonga_ngomo_ethnicity_2016}
Louppe, G., Al-Natsheh, H.~T., Susik, M., \& Maguire, E.~J. 2016, in Ethnicity {Sensitive} {Author} {Disambiguation} {Using} {Semi}-supervised {Learning}, ed. A.-C. Ngonga~Ngomo \& P.~Křemen, Vol. 649 (Cham: Springer International Publishing), 272--287, \dodoi{10.1007/978-3-319-45880-9_21}

\bibitem[{Mihaljević \& Santamaría(2021)}]{mihaljevic_disambiguation_2021}
Mihaljević, H., \& Santamaría, L. 2021, Scientometrics, 126, 3893, \dodoi{10.1007/s11192-021-03951-w}

\bibitem[{Mikolov {et~al.}(2013)Mikolov, Chen, Corrado, \& Dean}]{mikolov_efficient_2013}
Mikolov, T., Chen, K., Corrado, G., \& Dean, J. 2013, in Efficient {Estimation} of {Word} {Representations} in {Vector} {Space}.
\newblock \url{https://arxiv.org/pdf/1301.3781}

\bibitem[{Milojević(2013)}]{milojevic_accuracy_2013}
Milojević, S. 2013, Journal of Informetrics, 7, 767, \dodoi{10.1016/j.joi.2013.06.006}

\bibitem[{Montenegro(2023)}]{Montenegro_2023}
Montenegro, A. 2023, ORCID Public Data File 2023,  ORCID, \dodoi{10.23640/07243.24204912.v1}

\bibitem[{Müller {et~al.}(2017)Müller, Reitz, \& Roy}]{muller_data_2017}
Müller, M.-C., Reitz, F., \& Roy, N. 2017, Scientometrics, 111, 1467, \dodoi{10.1007/s11192-017-2363-5}

\bibitem[{Newman(2001)}]{newman_2001}
Newman, M. E.~J. 2001, Proceedings of the National Academy of Sciences, 98, 404, \dodoi{10.1073/pnas.98.2.404}

\bibitem[{Oord {et~al.}(2018)Oord, Li, \& Vinyals}]{oord_representation_2018}
Oord, A. v.~d., Li, Y., \& Vinyals, O. 2018.
\newblock \url{https://arxiv.org/pdf/1807.03748}

\bibitem[{Pal \& Savvides(2017)}]{Pal2017CopernicanL}
Pal, D.~K., \& Savvides, M. 2017, in Copernican loss : Learning a Discriminative Cosine Embedding.
\newblock \url{https://api.semanticscholar.org/CorpusID:201697921}

\bibitem[{Pooja {et~al.}(2022)Pooja, Mondal, \& Chandra}]{pooja_exploiting_2022}
Pooja, K., Mondal, S., \& Chandra, J. 2022, ACM Trans. Knowl. Discov. Data, 16, 88:1, \dodoi{10.1145/3502730}

\bibitem[{Portenoy {et~al.}(2022)Portenoy, Radensky, West, Horvitz, Weld, \& Hope}]{portenoy_bursting_2022}
Portenoy, J., Radensky, M., West, J.~D., {et~al.} 2022, in {CHI} {Conference} on {Human} {Factors} in {Computing} {Systems} (New Orleans LA USA: ACM), 1--13, \dodoi{10.1145/3491102.3501905}

\bibitem[{Priem {et~al.}(2022)Priem, Piwowar, \& Orr}]{priem_openalex_2022}
Priem, J., Piwowar, H., \& Orr, R. 2022, {OpenAlex}: {A} fully-open index of scholarly works, authors, venues, institutions, and concepts,  arXiv, \dodoi{10.48550/arXiv.2205.01833}

\bibitem[{Qian {et~al.}(2015)Qian, Zheng, Sakai, Ye, \& Liu}]{noauthor_dynamic_nodate}
Qian, Y., Zheng, Q., Sakai, T., Ye, J., \& Liu, J. 2015, Information Retrieval Journal, 18, 379

\bibitem[{Qiao {et~al.}(2019)Qiao, Du, Fu, Wang, \& Zhou}]{qiao_unsupervised_2019}
Qiao, Z., Du, Y., Fu, Y., Wang, P., \& Zhou, Y. 2019, in 2019 {IEEE} {International} {Conference} on {Big} {Data} ({Big} {Data}) (Los Angeles, CA, USA: IEEE), 910--919, \dodoi{10.1109/BigData47090.2019.9005458}

\bibitem[{Santini {et~al.}(2022)Santini, Gesese, Peroni, Gangemi, Sack, \& Alam}]{santini_knowledge_2022}
Santini, C., Gesese, G.~A., Peroni, S., {et~al.} 2022, Scientometrics, 127, 4887, \dodoi{10.1007/s11192-022-04426-2}

\bibitem[{Schneegans {et~al.}(2021)Schneegans, Lewis, \& Straza}]{schneegans_unesco_2021}
Schneegans, S., Lewis, J., \& Straza, T. 2021, {UNESCO} {Science} {Report}: the race against time for smarter development; executive summary - {UNESCO} {Digital} {Library}.
\newblock \url{https://unesdoc.unesco.org/ark:/48223/pf0000377250}

\bibitem[{Schneider {et~al.}(2023)Schneider, Alexander, \& Thomas}]{schneider_publications_2023}
Schneider, B., Alexander, J., \& Thomas, P. 2023, Publications {Output}: {U}.{S}. {Trends} and {International} {Comparisons} {\textbar} {NSF} - {National} {Science} {Foundation}.
\newblock \url{https://ncses.nsf.gov/pubs/nsb202333}

\bibitem[{Song {et~al.}(2015)Song, Kim, \& Kim}]{song_exploring_2015}
Song, M., Kim, E. H.-J., \& Kim, H.~J. 2015, Journal of Informetrics, 9, 924, \dodoi{10.1016/j.joi.2015.08.004}

\bibitem[{Strolger {et~al.}(2023)Strolger, Pegues, King, Miles, Ramsahoye, II, Blacker, \& Reid}]{strolger_pacman2_2023}
Strolger, L.-G., Pegues, J., King, T., {et~al.} 2023, The Astronomical Journal, 165, 215, \dodoi{10.3847/1538-3881/acc2c4}

\bibitem[{Strotmann {et~al.}(2010)Strotmann, Zhao, \& Bubela}]{strotmann_author_2010}
Strotmann, A., Zhao, D., \& Bubela, T. 2010, Author name disambiguation for collaboration network analysis and visualization - {Strotmann} - 2009 - {Proceedings} of the {American} {Society} for {Information} {Science} and {Technology} - {Wiley} {Online} {Library}.
\newblock \url{https://asistdl.onlinelibrary.wiley.com/doi/full/10.1002/meet.2009.1450460218}

\bibitem[{Subramanian {et~al.}(2021)Subramanian, King, Downey, \& Feldman}]{subramanian_s2and_2021}
Subramanian, S., King, D., Downey, D., \& Feldman, S. 2021, 2021 ACM/IEEE Joint Conference on Digital Libraries (JCDL), 170, \dodoi{10.1109/JCDL52503.2021.00029}

\bibitem[{Torvik \& Smalheiser(2009)}]{torvik_author_2009}
Torvik, V.~I., \& Smalheiser, N.~R. 2009, ACM transactions on knowledge discovery from data, 3, 11.
\newblock \url{https://pmc.ncbi.nlm.nih.gov/articles/PMC2805000/}

\bibitem[{Tran~H.(2014)}]{tran_author_2014}
Tran~H., Huynh~T., D.~T. 2014, 8397, 123, \dodoi{10.1007/978-3-319-05476-6_13}

\bibitem[{Varadharajalu {et~al.}(2011)Varadharajalu, Liu, \& Wong}]{wang_author_2011}
Varadharajalu, A., Liu, W., \& Wong, W. 2011, in {AI} 2011: {Advances} in {Artificial} {Intelligence}, ed. D.~Wang \& M.~Reynolds, Vol. 7106 (Berlin, Heidelberg: Springer Berlin Heidelberg), 152--161, \dodoi{10.1007/978-3-642-25832-9_16}

\bibitem[{Vishnyakova {et~al.}(2019)Vishnyakova, Rodriguez-Esteban, \& Rinaldi}]{vishnyakova_new_2019}
Vishnyakova, D., Rodriguez-Esteban, R., \& Rinaldi, F. 2019, Journal of the American Medical Informatics Association: JAMIA, 26, 1037, \dodoi{10.1093/jamia/ocz028}

\bibitem[{Wang {et~al.}(2011)Wang, Tang, Cheng, \& Yu}]{wang_adana_2011}
Wang, X., Tang, J., Cheng, H., \& Yu, P.~S. 2011, in 2011 {IEEE} 11th {International} {Conference} on {Data} {Mining} (Vancouver, BC, Canada: IEEE), 794--803, \dodoi{10.1109/ICDM.2011.19}

\bibitem[{Xiong {et~al.}(2021)Xiong, Bao, \& Wu}]{xiong_learning_2021}
Xiong, B., Bao, P., \& Wu, Y. 2021, Neural Computing and Applications, 33, 1987, \dodoi{10.1007/s00521-020-05088-y}

\bibitem[{Yan {et~al.}(2023)Yan, Odom, \& Natarajan}]{yan_knowledge-based_2023}
Yan, S., Odom, P., \& Natarajan, S. 2023, Knowledge-based {Refinement} of {Scientific} {Publication} {Knowledge} {Graphs},  arXiv, \dodoi{10.48550/arXiv.2309.05681}

\bibitem[{Zhang {et~al.}(2023)Zhang, Lu, \& Yang}]{zhang_span_2023}
Zhang, L., Lu, W., \& Yang, J. 2023, Journal of the Association for Information Science and Technology, 74, 168, \dodoi{10.1002/asi.24720}

\bibitem[{Zhang {et~al.}(2018)Zhang, Zhang, Yao, \& Tang}]{zhang_name_2018}
Zhang, Y., Zhang, F., Yao, P., \& Tang, J. 2018, in Proceedings of the 24th {ACM} {SIGKDD} {International} {Conference} on {Knowledge} {Discovery} \& {Data} {Mining}, {KDD} '18 (New York, NY, USA: Association for Computing Machinery), 1002--1011, \dodoi{10.1145/3219819.3219859}

\bibitem[{Zhou \& Sun(2024)}]{zhou_evaluating_2024}
Zhou, H., \& Sun, M. 2024, Evaluating authorship disambiguation quality through anomaly analysis on researchers' career transition,  arXiv, \dodoi{10.48550/arXiv.2412.18757}

\end{thebibliography}
\bibliographystyle{aasjournal} 

\appendix \label{sec:appendix}
\begin{table*}[t!]
\centering
\caption{LSPO dataset statistics and coverage}
\label{tab:metadata_comparison}
\begin{tabular}{lcccc}
\toprule
\textbf{Attribute} & \textbf{Total} & \textbf{Unique} & \textbf{Coverage} & \textbf{Description} \\
\midrule
Title & 553,496 & 442,840 & 100\% & Publication content \\
Abstract & 553,496 & 447,800 & 100\% & Publication content \\
Author Name & 553,496 & 159,660 & 100\% & Researcher information \\
Publication Venue & 553,496 & 8,293 & 100\% & Journal or Conference \\
Affiliation & 553,496 & 329,311 & 89.37\% & Researcher institution \\
ORCiD Identifier & 553,496 & 125,486 & 100\% & Label for training \\
Name block & 553,496 & 97,936 & 100\% & Grouping for disambiguation samples \\
\bottomrule
\end{tabular}%
\end{table*}

\section{Dataset Statistics Figure}
In Table~\ref{tab:metadata_comparison} we present the LSPO dataset statistics and coverage across the metadata utilized in training the NAND algorithm.
\begin{table*}[t!]
\centering
\caption{Results of AND on LSPO Test Set Across Various Representation Methods on an Untrained Network.}
\label{tab:notrain}
\begin{tabular}{l c c c c}
\toprule
\textbf{Representation} & \textbf{Acc.} & \textbf{Prec.} & \textbf{Recall} & \textbf{F1-score} \\
\midrule
Chars2vec + \textsc{Specter} & 85.90\,\% & 93.15\,\% & 85.03\,\% & 88.91\,\% \\
Chars2vec + Llama 3.2 & 75.96\,\% & 88.22\,\% & 73.65\,\% & 80.28\,\% \\
Word2vec + \textsc{Specter} & 66.46\,\% & 66.46\,\% & 100.00\,\% & 79.85\,\% \\
Word2vec + Llama 3.2 & 72.92\,\% & 81.36\,\% & 76.86\,\% & 79.05\,\% \\
\bottomrule
\end{tabular}%
\end{table*}%

\section{Software and Hardware}\label{app:hardware_software}
\subsection{Software}
We utilized the following software tools in our experiments:

\begin{enumerate}
    \item Python version 3.11.5\footnote{\url{https://www.python.org/}} for creating data pairs from the dataset.
    \item HuggingFace version 4.33.2\footnote{\url{https://huggingface.co/}}, Chars2Vec \footnote{\url{https://github.com/IntuitionEngineeringTeam/chars2vec}}, and Gensim\footnote{\url{https://radimrehurek.com/gensim/auto_examples/tutorials/run_word2vec.html}} for generating embeddings for different representation methods.
    \item Optuna version 3.6.0\footnote{\url{https://optuna.org/}} for hyperparameter tuning.
    \item PyTorch Lightning version 2.0.8 \footnote{\url{https://www.pytorchlightning.ai/}} for training the NAND model.
    \item PyTorch version 2.0.1\footnote{\url{https://pytorch.org/}} for training the NAND model. \citep{imambi2021pytorch}
\end{enumerate}

\subsection{Hardware}
Our experiments were performed on the High-Performance Computing Center (HPCC) at the Institute for Cyber-Enabled Research at Michigan State University. We estimate the following hardware usage:

\begin{enumerate}
    \item 48 hours of A100 GPU for generating embeddings from 2 transformer-based models.
    \item 190 hours of a combination of A100 and V100 GPUs for training our NAND models.
\end{enumerate}

\section{Preprocessing}\label{app:processing}
The following query was employed when matching titles from NASA/ADS to ORCiD profiles to avoid the linkage of non-unique titles (e.g., \verb|"Letter to the editors"| or \verb|"news and notes"|) to NASA/ADS authors, as these can be found across various fields in ORCiD, not only physics and astronomy.

\textbf{'len\_title > 2 and title != "additions and corrections" and title != "letter to the editor" and title != "news and notes" and title != "letters to the editor" and title != "Letter to the editor" and title != "Letter to the Editor" and title != "Letters to the Editor" and title != "Letters to the editor" and title != "From the editors" and title != "Response to the Letter to the Editor" and title != "List of contributors" and title != "List of Contributors"}

\section{Dataset and Code}\label{app:Code}
The LSPO dataset presented in this work is openly available at \url{https://zenodo.org/records/11489161}, and the code for training and evaluating the NAND model is available at \url{https://github.com/deepthought-initiative/neural_name_dismabiguator}. The repository includes:

\begin{itemize}
\item \textbf{AND\_dataset\_builder.py} — Generates training and evaluation pairs for the NAND models.
\item \textbf{AND\_nn\_exp.py} — Defines the model architecture, including the \textsc{Specter} + Chars2Vec configuration.
\item \textbf{AND\_readdata\_exp.py} — Data module for loading and preparing data for the NAND model.
\item \textbf{experiment.py} — Runs model training across multiple random seeds to measure performance variability.
\item \textbf{results.py} — Loads trained models and produces evaluation metrics.
\end{itemize}

To reproduce our results, clone the repository and follow the experiment workflow described in Section~\ref{sec:experiment}.

\begin{table*}[t!]
\centering
\caption{Hyperparameters for Models with Various Representation Methods, Loss Functions, Number of Outputs, and Batch Size}
\label{tab:hyperparams}
\setlength{\tabcolsep}{4pt}
\resizebox{\textwidth}{!}{%
\begin{tabular}{l l c c c c c c c c}
\toprule
\textbf{Representation} & \textbf{Loss} & \textbf{Layer} & \textbf{\# of} & \textbf{Activation} & \textbf{Learning} & \textbf{Dropout} & \textbf{Seeds} & \textbf{\# of} & \textbf{Batch} \\
\textbf{Methods} & \textbf{Functions} & \textbf{Size} & \textbf{Layers} & & \textbf{Rate} & & & \textbf{Outputs} & \textbf{Size} \\
\midrule
Chars2vec + \textsc{Specter} & Cosine-embedding & 2048 & 3 & SELU & 0.0001 & 0.2 & 1-5 & 128 & 2048 \\
Chars2vec + \textsc{Specter} & InfoNCE & 1024 & 1 & SELU & 0.00001 & 0.2 & 1-5 & 256 & 2048 \\
Chars2vec + Llama 3.2 & Cosine-embedding & 1024 & 4 & SELU & 0.00001 & 0.1 & 1-5 & 128 & 4096 \\
Chars2vec + Llama 3.2 & InfoNCE & 512 & 1 & SELU & 0.0001 & 0.3 & 1-5 & 128 & 2048 \\
Word2vec + \textsc{Specter} & Cosine-embedding & 512 & 3 & LeakyReLU & 0.00001 & 0.3 & 1-5 & 128 & 4096 \\
Word2vec + \textsc{Specter} & InfoNCE & 1536 & 1 & LeakyReLU & 0.00001 & 0.4 & 1-5 & 512 & 2048 \\
Word2vec + Llama 3.2 & Cosine-embedding & 512 & 1 & SELU & 0.001 & 0.1 & 1-5 & 512 & 50000 \\
Word2vec + Llama 3.2 & InfoNCE & 1024 & 1 & SELU & 0.0001 & 0.3 & 1-5 & 128 & 2048 \\
\bottomrule
\end{tabular}%
}
\end{table*}

\end{document}